# The spin exchange interaction effect on $T_c$ equation of anisotropic impure superconductors

P.Udomsamuthirun[(1)]* and R.Supadanaison [(1)]

[(1)] Department of Physics, Faculty of Science, Srinakharinwirot University, Bangkok 10110,Thailand.*E-mail: udomsamut55@yahoo.com

**Abstract**

We study the influence of spin exchange interaction of impurity scattering on critical temperature of anisotropic impure superconductors. The model of random nonmagnetic and magnetic impurity are revised to cover the effect of spin exchange interaction . The sign of magnitude of the second-order Born scattering have been changed after consideration the spin exchange interaction effect that also effect to form of $T_c$ equation. We can get the general $T_c$ equation that can be described anisotropic impure superconductors well and cover all model done before.

## 1.Introduction

The results of impurities that introduced into the superconductors are found in a change of the superconducting critical temperature . They modify the quasi-particle spectrum, interaction parameters and induce pair breaking in superconducting state. The non-magnetic impurities have little effect on critical temperature in s-wave superconductors [1] but they exhibit a strong pair breaking effect in the high temperature superconductors [2] . Shi and Li[3] study the influence of nonmagnetic impurities on critical temperature of the layered superconductors by using an superconducting-normal layer model with d-wave pairing in the superconducting layer. Haran and Nagi [4-7] proposed a theory of nonmagnetic and magnetic impurities in an anisotropic superconductors including the effect of anisotropic (momentum-dependent) impurities scattering. They consider an anisotropic superconductor with randomly distributing impurities, treating the electron-impurity scattering within second Born approximation, and neglecting the impurity-impurity interaction.  Although Haran and Nagi consider the effect of anisotropic  (momentum-dependent) impurities scattering, they  do not consider the effect of spin exchange interaction of impurities on $T_c$ equation . They get two kinds of $T_c$ equation. Firstly they find $T_c$ equation [5-7] that similar to Openov[8-9] and secondly they find $T_c$ equation with having the extra term[4].

Openov[8,9] study the effect of impurities, both magnetic and nonmagnetic, on the critical  temperature of anisotropy superconductors  as function of potential and spin-flip scattering rate. But his equation do not include  the effect of momentum-dependent of scattering potential and spin exchange interaction. Recently, Haran [10] also revise the model of anisotropic impurity superconductor by giving the condition of non-negative magnitude of the second-order Born scattering potential.

The previous work of Haran and Nagi [4-7]  and Opeov[8-9] have been described the effect of nonmagnetic and magnetic scattering on anisotropic superconductor but all of them show theirs  $T_c$ equations that having extra term in [4] and difference in some term . The physics behind this effect is not clear . In this work, we revise these works by considering  the effect of  spin exchange interaction of impurity scattering on $T_c$ equation that we think that it is the physics behind .Our model contain both non-negative and negative  magnitude of the second-order Born scattering potential that do not been considered before.After our calculation, we can get the only one $T_c$ equation that can describe all of the $T_c$ equation of [4-9], especially we can describe the extra term of [4].

## 2. Model and Calculation

Within the framework of BCS model, the Hamiltonian of a superconductor containing both nonmagnetic and magnetic impurities is as follows

$$H = \sum_{k,\sigma} \varepsilon_k a_{k\sigma}^+ a_{k\sigma} + \sum_{k,k',\sigma,\sigma'} u(k,\sigma;k',\sigma') a_{k\sigma}^+ a_{k'\sigma'} + \sum_{k,k'} V(k,k') a_{k\uparrow}^+ a_{-k\downarrow}^+ a_{-k'\downarrow} a_{k'\uparrow} \quad (1)$$

Where the operator $a_{k\sigma}^+ (a_{k\sigma})$ creates(annihilates) an electron with the wave vector $k$ and the spin projection on z-axis $\sigma = \uparrow$ or $\downarrow$, $\varepsilon_k = \xi_k - \mu$ is the (spin independent) quasiparticle energy measured from the chemical potential $\mu$, $u(k,\sigma;k',\sigma')$ is the matrix element for electron scattering by randomly distributed impurities(defects) from the state $(k',\sigma')$ to the state $(k,\sigma)$, and $V(k,k')$ is the BCS pair potential. $V(k,k') = -V_0 e(k) e(k')$ where $V_0$ is the pairing energy and $e(k)$ is a real basis function.

For the formal procedure of $T_c$ calculation, we define the normal and anomalous temperature Green's function as

$$G(k,\tau) = -<T_\tau a_{k\sigma}(\tau) a_{k\sigma}^+(0)>, \quad F(k,\tau) = <T_\tau a_{-k-\sigma}^+(\tau) a_{k\sigma}^+(0)>,$$

$$\tilde{G}(k,\tau) = -<T_\tau a_{-k-\sigma}^+(\tau) a_{-k-\sigma}(0)>, \quad \tilde{F}(k,\tau) = <T_\tau a_{k\sigma}(\tau) a_{-k-\sigma}(0)>$$

The matrix Green function in the Nambu representation is

$$\hat{G}(\omega,k) = \begin{pmatrix} G(\omega,k) & -\tilde{F}(\omega,k) \\ -F(\omega,k) & \tilde{G}(\omega,k) \end{pmatrix}$$

We can get the normal and anomalous temperature Green's function averaged over the impurity position and spin directions as

$$G(\omega,k) = -\frac{i\tilde{\omega} + \varepsilon_k}{\tilde{\omega}^2 + \varepsilon_k^2 + |\tilde{\Delta}(k)|^2} \quad (2)$$

$$F(\omega,k) = \frac{\tilde{\Delta}(k)}{\tilde{\omega}^2 + \varepsilon_k^2 + |\tilde{\Delta}(k)|^2} \quad (3)$$

where $\tilde{\omega}(k)$ is the renormalized Matsubara frequency and $\tilde{\Delta}(k)$ is the renormalized order parameter.

Although the isotropic impurity scattering can describe the impurity effect on superconductivity well, the weak anisotropic impurity scattering occur when the Fermi surface average of the order parameter is non zero . We can go beyond the isotropic scattering cases by treating a small perturbation of the isotropic scattering as [4-7]. The anisotropy impurity scattering can be found in case of extended magnetic moment impurity ( magnetic impurity scattering) and the anisotropy of the crystal lattice(nonmagnetic impurity scattering).

The momentum-dependent impurity potential can be simplify into model as

$$u(k,k') = v(k,k') + J(k,k')\vec{S} \cdot \vec{\sigma} \qquad (4)$$

where $v(k,k')$ is the spin-independent potential component and $J(k,k')\vec{S} \cdot \vec{\sigma}$ is the spin exchange interaction. $\vec{S}$ is the spin of magnetic impurity and $\vec{\sigma}$ is the electron spin density. We assumed a separable form of scattering probabilities.

$$v^2(k,k') = v_0^2 + v_1^2 f(k) f(k') \qquad (5)$$

$$J^2(k,k') = J_0^2 + J_1^2 g(k) g(k') \qquad (6)$$

where $v_0(v_1), J_0(J_1)$ are isotropic (anisotropic) scattering amplitudes for non-magnetic and magnetic potential. $f(k), g(k)$ are the momentum-dependent anisotropy function in the nonmagnetic and magnetic scattering channel .The averaged over the Fermi surface of $f(k)$ and $g(k)$ vanish , $< f(k) >=< g(k) >= 0$ , and are normalized as $< f^2(k) >= 1$ and $< g^2(k) >= 1$.

To consider the effect of impurity on anisotropic superconductor ,we set the equation [8,9]

$$\hat{G}^{-1}(\omega,k) = \hat{G}_0^{-1} - \hat{M}(\omega,k)$$

where $\hat{M}(\omega,k) =< u(k,k')\hat{G}(\omega,k')u(k',k) >$.

Finally we can get the $\tilde{\omega}(k)$ and $\tilde{\Delta}(k)$ [11,12] as

$$\tilde{\omega}(k) = \omega + \pi n_i N_0 \int_{FS} dS_{k'} n(k') \tilde{\omega}(k') \frac{v^2(k,k') + J^2(k,k')S(S+1)}{\sqrt{\tilde{\omega}^2(k') + \tilde{\Delta}^2(k')}} \qquad (7)$$

$$\tilde{\Delta}(k) = \Delta + \pi n_i N_0 \int_{FS} dS_{k'} n(k') \tilde{\Delta}(k') \frac{v^2(k,k') \pm J^2(k,k')S(S+1)}{\sqrt{\tilde{\omega}^2(k') + \tilde{\Delta}^2(k')}} \qquad (8)$$

Here $\omega = \pi T(2n+1)$ ,T is temperature ,$n$ is integer number. $n_i$ is impurity concentration.

The "$\pm$" sign in Eq.(8) play an important rule in this model. They depend on the spin exchange interaction of magnetic impurity scattering that effect to the sign of magnitude of the second-order Born scattering. The effect of magnetic impurities scattering contribution from both spin-flip and spin-conserving scattering of electrons due to their exchange interaction with magnetic impurities.

The "+" and "-" sign mean that we consider the effect of non-negative and negative magnitude of the second-order Born scattering, respectively. After consideration the spin exchange interaction term Eq.(4), we find that "+"sign can be occurred in case of spin-conserving scattering that $J(k,k')\vec{S}\cdot\vec{\sigma}$ is equal to $J(k,k')\vec{S}\cdot\vec{\sigma} \equiv J(k,k')S^z$ and spin-flip scattering as $J(k,k')(-i\gamma_\sigma)S^x$ or $J(k,k')S^y$. And the "-" sign[8-9] can be found in case spin-conserving scattering that $J(k,k')\vec{S}\cdot\vec{\sigma}$ is equal to $J(k,k')\vec{S}\cdot\vec{\sigma} \equiv J(k,k')\gamma_\sigma S^z$ and spin-flip scattering as $J(k,k')S^x$ or $J(k,k')i\gamma_\sigma S^y$. Where $\gamma_\sigma = +1$ and $-1$ for $\sigma = \uparrow$ and $\downarrow$ respectively, and $S^x, S^y, S^z$ are the Pauli matrices of spin of magnetic impurity. The "$\pm$" sign in Eq.(8) show the orientation of electron spin through $\gamma_\sigma$ contributed from spin-conserving exchange scattering of electrons due to their exchange interaction with magnetic impurities.

Taking a separable pair potential $V(k,k') = -V_0 e(k)e(k')$, that $V_0$ is the pairing energy and $\Delta(k)$ is the orbital part of the singlet superconducting order parameter. We defined as $\Delta(k) = \Delta\, e(k)$ where $e(k)$ is a real basis function and $<e^2> = 1$, where $<..>$ denotes the average value over the Fermi surface that $\int \frac{d^3k}{(2\pi)^3} \to N_0 \int_{FS} dS_k n(k) \int d\xi_k$ and $<..> = \int_{FS} dS_k n(k)(..)$. Here $n(k)$ is the angle-resolved Fermi surface density of state. $N_0$ is the overall density of state at Fermi surface.

We can get the critical temperature as

$$\ln(\frac{T_C}{T_{C0}}) = (1-<e>^2)[\Psi\left(\frac{1}{2}\right) - \Psi\left(\frac{1}{2} + (\frac{\Gamma_0 + G_0}{2\pi T_C})\right)] + S_1 + S_2 + S_3 \qquad (9)$$

That

$$S_1 = 2\pi T_c \sum_{\omega>0} \frac{<ef>}{(\omega+\Gamma_0+G_0)} \frac{\Gamma_1<ef>(\omega+\Gamma_0+G_0 \mp G_1)\pm <fg><eg>G_1\Gamma_1}{(\omega+\Gamma_0+G_0-\Gamma_1)(\omega+\Gamma_0+G_0 \mp G_1)\mp <fg>^2 \Gamma_1 G_1}$$

$$S_2 = \pm 2\pi T_c \sum_{\omega>0} \frac{<eg>}{(\omega+\Gamma_0+G_0)} \frac{G_1<eg>(\omega+\Gamma_0+G_0-\Gamma_1)+<fg><ef>G_1\Gamma_1}{(\omega+\Gamma_0+G_0 \mp G_1)(\omega+\Gamma_0+G_0-\Gamma_1)\mp <fg>^2 \Gamma_1 G_1}$$

$$S_3 = <e>^2 [\Psi(\frac{1}{2}) - \Psi(\frac{1}{2}+\frac{G_0 \pm G_0}{2\pi T_c})]$$

Here $\Gamma_0 = \pi n_i N_0 v_0^2$, $\Gamma_1 = \pi n_i N_0 v_1^2$, $G_0 = \pi n_i N_0 J_0^2 S(S+1)$, $G_1 = \pi n_i N_0 J_1^2 S(S+1)$ are scattering rate of isotropic non-magnetic, anisotropic non-magnetic, isotropic magnetic, and anisotropic magnetic channel, respectively. $G_0$ and $G_1$ depend on the magnitude of spin exchange interaction.

The Eq.(9) is the critical temperature equation of anisotropic impure superconductors that include the effect of random magnetic and non-magnetic impurities, and the spin exchange interaction.

### 3. Discussion

In our model, we can get the $T_c$ equation that can describe all case done before[4-7,9,10]. The effect of spin exchange interaction are to change the sign of magnitude of the second-order Born scattering shown by Eq.(8). And they effect to the critical temperature equation by changing sign from plus to minus of Eq.(9). The occurring of "$\pm$" sign in Eq.(9) (in $S_1, S_2$ and $S_3$) depend on the sign "$\pm$" of Eq.(8). In case of "-" sign, we can get the result of Haran and Nagi[4] especially

$S_3 = <e>^2 [\Psi(\frac{1}{2}) - \Psi(\frac{1}{2}+\frac{2G_0}{2\pi T_c})]$ that they think it is the additional term. Within our

model, we can clarify that this additional term come from the spin exchange interaction. In case of "+" sign, we can get the results of Haran and Nagi[5-7], and Openov[8,9]. The term $S_3$ equal to zero in this case. The new term of $S_1$ and $S_2$ have been found by our calculation.

For above discussion we can conclude that the spin exchange interaction can change the form of $T_c$ equation and the final $T_c$ equation can describe the critical temperature of anisotropic impure superconductors well. In this work we will not shown the numerical calculation because they are the special case of our consideration that have been done before by the others .

## 4. Conclusion

We have revised the model of nonmagnetic and magnetic impurity of Haran and Nagi[4-7] and Opeov [8,9] that consider the problem of nonmagnetic and magnetic impurities in an anisotropic impure superconductor for the case of anisotropic(momentum-dependent) impurity scattering in weak-coupling approximation. The effect of spin exchange interaction of impurity scattering on $T_c$ equation are included in our model. The spin exchange interaction effect to the sign of magnitude of the second-order Born scattering that also effect to $T_c$ equation. After our calculation, we can get the general $T_c$ equation that can describe every model done before well .

**Acknowledgement**

The author would like to thank Professor Dr.Suthat Yoksan for the useful discussion and also thank Thai Research Fund and Office of Higher Education Commission for the financial support.